\documentclass[sigconf]{acmart}

\usepackage{multirow}
\usepackage{balance}
\usepackage{listings}
\usepackage{enumitem}
\usepackage[most]{tcolorbox}

\lstdefinestyle{python_style}{
    basicstyle=\small\ttfamily,
    breakatwhitespace=false,    
    breaklines=true,         
    captionpos=b,               
    keepspaces=true,        
    numbers=left,         
    numbersep=5pt,
    showspaces=false,        
    showstringspaces=false,
    showtabs=false,            
    tabsize=2
}
\newcommand{\framedtext}[1]{%
\par%
\noindent\fbox{%
    \parbox{\dimexpr\linewidth-2\fboxsep-2\fboxrule}{#1}%
}%
}

 \AtBeginDocument{%
 }

\copyrightyear{2023}
\acmYear{2023}
\setcopyright{rightsretained}
\acmConference[SIGIR '23]{Proceedings of the 46th International ACM SIGIR Conference on Research and Development in Information Retrieval}{July 23--27, 2023}{Taipei, Taiwan}
\acmBooktitle{Proceedings of the 46th International ACM SIGIR Conference on Research and Development in Information Retrieval (SIGIR '23), July 23--27, 2023, Taipei, Taiwan}\acmDOI{10.1145/3539618.3591887}
\acmISBN{978-1-4503-9408-6/23/07}

\makeatletter
\gdef\@copyrightpermission{
  \begin{minipage}{0.3\columnwidth}
   \href{https://creativecommons.org/licenses/by/4.0/}{\includegraphics[width=0.90\textwidth]{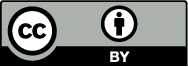}}
  \end{minipage}\hfill
  \begin{minipage}{0.7\columnwidth}
   \href{https://creativecommons.org/licenses/by/4.0/}{This work is licensed under a Creative Commons Attribution International 4.0 License.}
  \end{minipage}
  \vspace{5pt}
}
\makeatother

\begin{document}

\title{MMEAD: MS MARCO Entity Annotations and Disambiguations}

\author{Chris Kamphuis}
\email{ckamphuis@cs.ru.nl}
\affiliation{%
  \institution{Radboud University}
  \country{The Netherlands}
}

\author{Aileen Lin}
\email{l43lin@uwaterloo.ca}
\affiliation{%
  \institution{University of Waterloo}
  \country{Canada}
}

\author{Siwen Yang}
\email{s243yang@uwaterloo.ca}
\affiliation{%
  \institution{University of Waterloo}
  \country{Canada}
}

\author{Jimmy Lin}
\email{jimmylin@uwaterloo.ca}
\affiliation{%
  \institution{University of Waterloo}
  \country{Canada}
}

\author{Arjen P. de Vries}
\email{arjen@acm.org}
\affiliation{%
  \institution{Radboud University}
  \country{The Netherlands}
}

\author{Faegheh Hasibi}
\email{f.hasibi@cs.ru.nl}
\affiliation{%
  \institution{Radboud University} 
  \country{The Netherlands}
}

\renewcommand{\shortauthors}{Chris Kamphuis et al.}

\begin{abstract}
  MMEAD, or MS MARCO Entity Annotations and Disambiguations, is a resource for entity links for the MS MARCO datasets. We specify a format to store and share links for both document and passage collections of MS MARCO. Following this specification, we release entity links to Wikipedia for documents and passages in both MS MARCO collections (v1 and v2). Entity links have been produced by the REL and BLINK systems. 
  MMEAD is an easy-to-install Python package, allowing users to load the link data and entity embeddings effortlessly. Using MMEAD takes only a few lines of code. Finally, we show how MMEAD can be used for IR research that uses entity information. We show how to improve recall@1000 and MRR@10 on more complex queries on the MS MARCO v1 passage dataset by using this resource. We also demonstrate how entity expansions can be used for interactive search applications. 
\end{abstract}



\maketitle

\section{Introduction}
The MS MARCO datasets~\citep{msmarco} have become the \emph{de facto} benchmark for evaluating deep learning methods for Information Retrieval (IR). The TREC deep learning track~\citep{trec-dl}, which has run since 2019, derives its datasets from the MS MARCO passage and document collections. The collections have been used in zero- and few-shot scenarios for diverse retrieval tasks and domains~\citep{thakur2021beir, thakur2022domain, xu2022laprador}. They also serve as primary resources for training deep learning models for downstream IR tasks such as conversational search~\citep{dalton2021cast} and search over knowledge graphs~\citep{Gerritse22} to achieve state-of-the-art results.

Purely text-based neural IR models, trained using MS MARCO collections, can generally not reason over complex concepts in the social and physical world~\citep{bosselut2021dynamic, sciavolino:2021:simple}. In response, recently proposed neuro-symbolic methods aim to combine neural models and symbolic AI approaches, e.g., by using knowledge graphs, which map concepts to symbols and relations. An essential step in developing neuro-symbolic models is connecting text to entities that represent the world's concepts formally. This step is mainly done using \textit{Entity linking}, an intermediary between text and knowledge graphs, which detects entity mentions in the text and links them to the corresponding entries in a knowledge graph.

Despite the proven effectiveness of neuro-symbolic AI -- and for IR models in particular~\citep{Tran:2022:DRE, Gerritse22, Chatterjee:2022:BERTER} -- the IR community has made limited efforts to develop such models. A primary hindrance is the annotation of large-scale collections with entities; entity linking methods are computationally expensive. Running them over a large text corpus (e.g., MS MARCO v2 with 12M documents and 140M passages) requires extensive resources. This paper aims to fill this gap by making entity annotations of the MS MARCO ranking collections readily available and easy to use.

With this work, we publish MMEAD,\footnote{MMEAD is pronounced as the drink mead.} a resource that provides entity links for the MS MARCO document and passage ranking collections. Two state-of-the-art entity linking tools, namely REL~\citep{rel, rebl} and BLINK~\citep{blink}, are utilized for annotating the corpora. The annotations are stored in a DuckDB database, enabling efficient analytical operations and fast access to the entities. The resource is available as a Python package and can be installed from PyPI effortlessly. The resource also includes a sample demo, enabling queries with complex compositional structures about entities. 

We envision that MMEAD will foster research in neuro-symbolic IR research and can be used to further improve neural retrieval models. In our experiments, we show significant improvements on recall for neural re-ranking IR models when using MMEAD annotations as bag-of-word expansions for queries and passages. Our experiments reveal that the difference in effectiveness is even greater (in terms of both recall and MRR) for complex queries that require further reasoning over entities.

To show the usefulness of our resource, we also present how to enrich interactive search applications. Specifically, we demonstrate how to obtain entities' geographical locations by relating the entities found in passages to their Wikidata entries. Plotting these entities on the world map shows that the MS MARCO passages can be geo-located all over the world.
We can also move from location to web text by retrieving all passages associated with a geographical location that we present through an interactive demo.  

\smallskip
\noindent In summary, this paper makes the following contributions:

 \begin{itemize}[leftmargin=*]
    \item We annotate the documents of the MS MARCO passage and document collections and share these annotations. By sharing these annotations, we ease future research in neuro-symbolic retrieval, which extensively uses entity information. We also provide useful metadata such as Wikipedia2Vec~\cite{wikipedia2vec} entity embeddings. 
    \item We provide a Python library that makes our data easy to use. All data is stored in DuckDB tables, which can be loaded and queried quickly. The library is easy to install through PyPI, and the entity annotations are available with only a few lines of code.
    \item We experimentally show that retrieval effectiveness measured by recall significantly increases when using MMEAD. The improvement is even greater for hard queries, where we observe low retrieval effectiveness using text-only IR models.
    \item We demonstrate how the data can be used in geographical applications. For example, we can plot on a static map all entities found in the MS MARCO v2 passage collection for which geographical data is available. Additionally, through an interactive demo, we can retrieve all passages associated with a geographical location.  
 \end{itemize}

\noindent MMEAD is publicly available at \url{https://github.com/informagi/mmead}.

\section{Background}

In this section, we describe systems that are used for creating entity annotations on the MS MARCO collections for MMEAD. 

\subsection{REL}
REL (Radboud Entity Linker)~\citep{rel} is a state-of-the-art open-source entity linking tool designed for high throughput and precision. REL links entities to a knowledge graph (Wikipedia) using a three-stage approach: (1) mention detection, (2) candidate selection, and (3) entity disambiguation. We briefly explain these three steps:

\begin{enumerate}[leftmargin=*]
    \item \emph{Mention Detection.} REL starts the entity linking process by first identifying all text spans that might refer to an entity. In this stage, it is essential that all possible entities in the text are identified, as only the output of this stage can be considered an entity by REL. These spans are identified using a named entity recognition (NER) model based on contextual word embeddings. For our experiments, we use the NER model based on Flair embeddings. 
    \item \emph{Candidate Selection.} Up to seven candidate entities are considered for every mention found by Flair. Part of these entities are selected according to the prior probability $P(e|m)$ of the mention $m$ being linked to the entity $e$. Precisely, the top-4 ranked entities based on $P(e|m) = \min(1, P_{\mathit{Wiki}}(e|m) + P_{\mathit{YAGO}}(e|m))$ are selected, where $P_{\mathit{YAGO}}(e|m))$ is a uniform probability from the YAGO dictionary~\citep{yago} and $P_{\mathit{Wiki}}(e|m)$ is computed based on the summation of hyperlink counts in Wikipedia and the CrossWikis corpus~\citep{crosswiki}.
    The remaining three candidate entities are determined according to the similarity of an entity and the context of a mention. For the top-ranked candidates based on $P(e|m)$ probabilities, the context similarity is calculated by $\mathbf{e}^T \sum_{w\in c}\mathbf{w}$. Here $\mathbf{e}$ is the entity embedding for entity $e$, and $\mathbf{w}$ are the word embeddings in context $c$, with a maximum length of 100-word tokens. The entity and word embeddings are jointly learned using Wikipedia2Vec~\citep{wikipedia2vec}. 
    \item \emph{Entity Disambiguation.} The final stage tries to select the correct entity from the candidate entities and maps it to the corresponding entry in a knowledge graph (Wikipedia). For this, REL assumes a latent relation between entities in the text and utilizes the Ment-norm method proposed by~\citet{ment-norm}.
\end{enumerate}

\noindent REL is designed to be a modular system, making it easy to swap, for example, the NER system with another. All necessary scripts to train the REL system are available on GitHub,\footnote{\url{https://github.com/informagi/rel}, last accessed April 26th 2023} making it easy to update REL to a more recent Wikipedia dump. Recently, a batch extension of REL, REBL~\citep{rebl}, was released, which improves the efficiency of REL for large-scale annotations, particularly in the candidate selection and entity disambiguation stages.  

\subsection{BLINK}
BLINK~\citep{blink} is a BERT-based~\citep{BERT} model for candidate selection and entity disambiguation, which assumes that entity mentions are already given. When utilized in an end-to-end entity linking setup, BLINK achieves similar effectiveness scores as REL. Below we describe the three steps of mention detection, candidate selection, and entity disambiguation for end-to-end entity linking using BLINK.

\begin{enumerate}[leftmargin=*]
    \item \emph{Mention Detection.} The mention detection stage can be done using an NER model. Like REL, we utilized Flair NER ~\citep{flair} for mention detection.
    \item \emph{Candidate Selection.} BLINK considers ten candidates for each mention. The candidates are selected through a bi-encoder (similar to~\citet{poly-encoders}) that embeds mention contexts and entity descriptions. The mention and the entity are encoded into separate vectors using the \texttt{[CLS]} token of BERT. The similarity score is then calculated using the dot-product of the two vectors representing the mention context and the entity.  
    \item \emph{Entity Disambiguation.} For entity disambiguation, BLINK employs a cross-encoder to re-rank the top 10 candidates selected by the candidate selection stage. The cross-encoder usage is similar to the work by~\citet{poly-encoders}, which employs a cross-attention mechanism between the mention context and entity descriptions. The input is the concatenation of the mention text and the candidate entity description.  
\end{enumerate}

\subsection{DuckDB}

DuckDB~\citep{duckdb} is an in-process column-oriented database management system. It is designed with requirements that are beneficial for the MMEAD resource:

\begin{enumerate}[leftmargin=*]
    \item \emph{Efficient analytics.} DuckDB is designed for analytical (OLAP) workloads, while many other database systems are optimized for transactional queries (OLTP). DuckDB is especially suitable for cases where analytics are more important than transactions. As we release a resource, transactions (after loading the data) are unnecessary, making an analytics database more useful than a transactional-focused one. 
    \item \emph{In-process.} DuckDB runs in-process, which means no database server is necessary, and all data processing happens in-process. This allows the database to be installed from PyPI without any additional steps. 
    \item \emph{Efficient data transfer.} Because DuckDB runs in-process, it can transfer data from and to the database more easily, as the address space is shared. In particular, DuckDB uses an API built around NumPy and Pandas, which makes data (almost) immediately available for further data analysis within Python. 
\end{enumerate}

\noindent DuckDB also supports the JSON and parquet file formats, making data loading especially fast when data is provided in such formats.

\section{MMEAD}
MMEAD provides links for MS MARCO collections v1 and v2 created by the REL entity linker, and links for the MS MARCO v1 passage collection by the BLINK entity linker. For REL, we use its batch entity linking extension, REBL~\citep{rebl}. The knowledge graphs used for the REL and BLINK entity linkers are Wikipedia dumps from 2019-07 and 2019-08, respectively. Both dumps are publicly available from the linking systems' Github pages. 

\subsection{Goals}
The design criteria for MMEAD are based on the following goals:

\begin{itemize}[leftmargin=*]
    \item \emph{Easy-to-use.} It should be easy to load and use the linked entities in experiments. With only a few lines of code, it should be possible to load entities and use them for analysis. Additional information should also be readily available, like where entities appear in the text and their latent representations.
    \item \emph{High-quality entity links.} We wish to release high-quality entity links for the MS MARCO collections, so that applying neuro-symbolic models and reasoning over entities becomes feasible.
    \item \emph{Extensibility.} It should be easy to link the collections with a different entity linking system and publish them in the same format as MMEAD. This way, we can integrate links produced by other entity linking systems and make them automatically available through the MMEAD framework.
    \item \emph{Useful metadata.} Additional data that can help with experiments should be provided; this includes mapping entities to their respective identifiers and latent representations. 
\end{itemize}

\subsection{Design}
\paragraph{Easy-to-use.} To create an easy-to-use package, we make the MMEAD data publicly available as JSONL files, which is the same format as the MS MARCO v2 collections. Each line of JSON contains entity links for one of the documents or passages in the collections; see Figure~\ref{fig:json-example-passage-v1}. The corresponding document can be identified through the JSON field that represents the document/passage identifier: \texttt{docid} for documents and \texttt{pid} for passages. Then, for every section of a document, a separate JSON field is available to access the entities in that section. For passages, there is only one section containing the entity annotations of the passage, while for MS MARCO v2 documents, we link not only the \texttt{body} of the document but also the \texttt{header} and the \texttt{title}.

All essential information about the entity mentions and linked entities is stored in the JSON objects. 
Specifically, the following metadata is made available: \texttt{entity\_id}, \texttt{start\_pos}, \texttt{end\_pos}, \texttt{entity}, and \texttt{details}. The field \texttt{entity\_id} stores the identifier that refers to the entry in the knowledge graph (Wikipedia, in our case). The \texttt{start\_pos} and \texttt{end\_pos} fields store the start and end positions of the text span that refers to the linked entity (i.e., as a standoff annotation of the entity mention). The positions are UTF-8 indices into the text, ready to be used in Python to extract the relevant parts of the document. The field \texttt{entity} stores the text representation of the entity from the knowledge graph. 
We chose to store this field for convenience and human readability. The \texttt{details} field is a JSON object that stores linker-specific information; examples include the entity type available from the NER module and the confidence of the identified mention.

\paragraph{High-quality entity links.} MMEAD provides entity links produced by state-of-the-art entity linking systems. For this paper, we provide links from REL for both MS MARCO v1 and v2 passages and docs, and links from BLINK for MS MARCO v1 passages. Both these systems have high precision, ensuring that identified mentions and their corresponding entities are likely correct. The knowledge graphs used by the entity linkers are the same as those used in the original studies; this way, extensive research has been done to confirm the precision of the linking systems.

\paragraph{Extensibility.} We ensure extensibility by clearly describing the format in which the entity links are provided. If another system shares its links in the same format, the MMEAD Python library can work with the data directly. The \texttt{details} field per entity annotation enables inclusion of linker-specific information. REL provides specific instructions on updating the system to newer versions of Wikipedia in its documentation, making it possible to easily release links to newer versions of Wikipedia.

\paragraph{Useful metadata.} Alongside the entity links, we also provide additional useful metadata. Specifically, we release Wikipedia2Vec~\citep{wikipedia2vec} embeddings (300d and 500d feature vectors). REL uses the 300d Wikipedia2Vec feature vectors internally for candidate selection. These feature vectors consist of word embeddings \emph{and} entity embeddings mapped into the same high-dimensional feature space. These embeddings can be used directly for information retrieval research~\citep{Gerritse:2020:GEER, Gerritse22}. We also release a mapping of entities to their identifiers. The entity descriptions can change in different versions of Wikipedia, but their identifiers remain constant.
The identifier can also be used to find the corresponding entity in other knowledge graphs such as Wikidata.

\subsection{An Example}

A passage from the MS MARCO v1 passage ranking collection is shown below.\footnote{This is the second passage from the collection.}

\smallskip
\begin{center}
  \fbox{
    \begin{minipage}{0.4\textwidth}
        \emph{The Manhattan Project and its atomic bomb helped bring an end to World War II. Its legacy of peaceful uses of atomic energy continues to have an impact on history and science.}
    \end{minipage}
    }    
\end{center}
\smallskip
A few text spans in this text can be considered as entities: ``the Manhattan Project'', ``World War II'', and ``atomic energy.'' REL identifies two of these entities: the \emph{Manhattan Project} and \emph{World War II}. 
The output of the system is converted to our JSON specification, which results in the JSON object presented in Figure~\ref{fig:json-example-passage-v1}. The value of the \texttt{md\_score} field shows that Flair is more certain about ``World War II'' being an entity than the ``Manhattan Project.'' 

\begin{figure}[t]
\begin{lstlisting}[frame=single, numbers=none]
{
    "passage": [
        {
            "entity_id": 19603, 
            "start_pos": 4, 
            "end_pos": 21,
            "entity": "Manhattan Project",
            "details": {
                "tag": "ORG",
                "md_score": 0.613243
            }
        }, 
        {
            "entity_id": 32927,
            "start_pos": 65,
            "end_pos": 77,
            "entity": "World War II",
            "details": {
                "tag": "MISC",
                "md_score": 0.991474
            }
        }
    ], 
    "pid": 1
}
\end{lstlisting}
\caption{Example of MMEAD annotations for a MS MARCO passage in JSON format. The field \texttt{tag} depicts the type of the entity and \texttt{md\_score} shows the certainty of the mention detection component in identifying the text span as a mention.}
\label{fig:json-example-passage-v1}
\end{figure}

Table~\ref{number-links} shows the number of entities found in the collections by the REL system. Blink found 21,968,356 entity links for the v1 passage collection.  For 11,177,904 entities, the two linking systems produced exactly the same output. 

\begin{table}[t]
    \centering
    \caption{Number of entities linked by REL; we show the total number of entities found and how many entities there are per passage/document on average.}
    \begin{tabular}{c|c|c}
    \toprule
    & passages & docs \\
    \midrule
    MS MARCO v1 & 18,561,221 (2.10) & 145,725,732 (45.34) \\
    MS MARCO v2 & 233,254,024 (1.69) & 661,183,287 (55.28) \\
    \bottomrule
    \end{tabular}
    \label{number-links}
\end{table}

\section{How To Use}

MMEAD comes with easy-to-use Python code, allowing users to work with the resource effortlessly. To start, MMEAD can be installed from PyPI using pip:

\begin{verbatim}
    $ pip install mmead
\end{verbatim}

\noindent After installation, the entity links can be loaded into a DuckDB~\citep{duckdb} database with only a couple of lines of code, as shown in Figure~\ref{fig:load-links}.
\begin{figure}[!t]
\begin{lstlisting}[language=python]
>>> from mmead import get_links
>>> links = get_links('v1', 'passage', linker='rel')
\end{lstlisting}
\caption{Example of how to load MMEAD entity links for the MS MARCO v1 passage collection.}
\label{fig:load-links}
\end{figure}
When running this code for the first time, initialization will take some time, as all the data need to be downloaded and ingested into the DuckDB database. After loading the data for the first time, it is automatically stored on disk. Loading the persisted data for later usage will only take seconds.

Once the data is loaded, it is ready to use. We provide a simple interface to access the data. The code shown in Figure~\ref{fig:load-links-for-document} loads the entity links available for a document in the MS MARCO v1 passage ranking collection. When using this function, the data is provided in JSON format, making it easy to access the annotations.

\begin{figure}[!t]
\begin{lstlisting}[language=python]
>>> links.load_links_from_docid(123)
{"passage":[{"entity_id":"7954681", ... }
\end{lstlisting}
\caption{Example of how to load the entity links for a document. For formatting reasons, we do not show the full output. }
\label{fig:load-links-for-document}
\end{figure}

We also provide word and entity embeddings generated by Wikipedia2Vec~\citep{wikipedia2vec} based on the 2019-07 Wikipedia dump. These embeddings are stored in DuckDB tables and are available as Numpy arrays after loading. Figure~\ref{fig:dot-product} shows how embeddings are loaded using MMEAD. The example demonstrates that the entity embedding of \emph{Montreal} and the word embedding of ``Montreal'' are closer to each other than the word embeddings of the two words ``Montreal'' and ``green'' based on dot-product as a similarity function. The dimensionality of the embedding vectors (300 or 500) can be specified in the code.

\begin{figure}[!t]
\begin{lstlisting}[language=python]
>>> from mmead import get_embeddings
>>> e = get_embeddings(300, verbose=False)
>>> montreal_word = \
    e.load_word_embedding("Montreal")
>>> montreal_entity = \
    e.load_entity_embedding("Montreal")
>>> green_word = \
    e.load_word_embedding("green")

>>> montreal_word @ montreal_entity
31.83191792
>>> montreal_word @ green_word
5.55568354

>>> toronto_word = \
    e.load_word_embedding("Toronto")
>>> toronto_word
array([-1.497e-01, -7.765e-01, -1.000e-02, ...])
>>> montreal_word @ toronto_word
21.62585146
\end{lstlisting}
\caption{Example code for loading word and entity embeddings. It shows that the dot-product between ``Montreal'' word and entity embeddings is greater than the dot-product of embedding vectors for the word ``Montreal'' and a random word. The word embeddings of Montreal and Toronto, two cities in Canada, are more similar.}
\label{fig:dot-product}
\end{figure}

The mapping between the official Wikipedia identifiers and entity text representations is extracted from the 2019-07 Wikipedia dump. If entity annotations from another version of Wikipedia are available, the MMEAD mappings can be used to match entities between the dumps. 
Needless to say, emerging entities in newer versions of Wikipedia cannot be mapped to the version that is available in MMEAD. However, existing entities in MMEAD can be mapped to newer versions of Wikipedia in a straightforward manner.
Figure~\ref{fig:load_mappings} shows how entity identifiers can be matched to their text and the other way around.  

\begin{figure}[!t]
\begin{lstlisting}[language=python]
>>> from mmead import get_mappings
>>> m = get_mappings(verbose=False)
>>> m.get_id_from_entity("Montreal")
7954681
>>> m.get_entity_from_id(7954681)
'Montreal'
\end{lstlisting}
\caption{Entity names and identifiers are accessible in MMEAD. Given an entity text, we can directly find its corresponding identifier and vice versa.}
\label{fig:load_mappings}
\end{figure}

As DuckDB is used as a database engine for MMEAD, it is possible to directly access the underlying tables and issue structured queries in an efficient manner. Figure~\ref{fig:sql_engine} shows an example, where a connection to the database is created, and the identifiers of passages containing the entity \emph{Nijmegen} are retrieved.

All data can be downloaded directly as well, and links to the data are provided on our Github page.\footnote{\url{https://github.com/informagi/mmead}}

\begin{figure}[!t]
\begin{lstlisting}[language=python]
>>> from mmead import load_links
>>> cursor = load_links(
...     'msmarco_v1_passage_links',
...     verbose=False
... )
>>> cursor.execute("""
...     SELECT pid 
...     FROM msmarco_v1_passage_links_rel 
...     WHERE entity='Nijmegen'
... """)
>>> cursor.fetchall()
[(771129,), (1273612,), (1418035,), ... ]
\end{lstlisting}
\caption{All data is stored in DuckDB tables, and thus it is possible to directly access the tables and issue queries. In this example, we extract the identifiers of passages that contain the city of Nijmegen.}
\label{fig:sql_engine}
\end{figure}

\section{Entity Expansion with MMEAD}

To demonstrate the usefulness of MMEAD for (neural) retrieval models, we have conducted experiments that extend existing models with MMEAD annotations.
These experiments serve a demonstrative purpose only, and the full potential of this resource is to be further explored in (neuro-)symbolic IR models~\citep{Gerritse22,Tran:2022:DRE}.

\begin{table*}[!t]
    \centering
    \caption{Results on the MS MARCO v1 passage collection, using only the queries that have entity annotations. Bolded numbers are the highest achieved effectiveness. Scores with a dagger (\dag) are significantly better compared to BM25 with no expansion (run \emph{a}), following a paired t-test with Bonferroni correction. For MRR, we have not calculated significance scores due to its ordinal scale~\citep{fuhr-mrr}.}
    \begin{tabular}{rl|llll|llll}
        \toprule
        \multirow{2}{*}{}
        & & \multicolumn{4}{c|}{R@1000} & \multicolumn{4}{c}{MRR@10} \\
        & & \multicolumn{1}{c}{dev} & \multicolumn{1}{c}{hard} & \multicolumn{1}{c}{harder} & \multicolumn{1}{c|}{hardest} & \multicolumn{1}{c}{dev} & \multicolumn{1}{c}{hard} & \multicolumn{1}{c}{harder} & \multicolumn{1}{c}{hardest} \\
        \midrule
        a. & BM25 -- No Expansion                           & 0.9111 & 0.7855 & 0.7444 & 0.6677 & \textbf{0.2413} & 0.0373 & 0.0137 & 0.0000 \\
        b. & BM25 -- Entity Text                            & 0.9183 & 0.8240\dag & 0.7951\dag & 0.7298\dag & 0.2202 & 0.0385 & 0.0173 & \textbf{0.0057} \\
        c. & BM25 -- Entity Hash                            & 0.9105 & 0.7980 & 0.7576 & 0.6848 & 0.2199 & 0.0383 & \textbf{0.0175} & 0.0052 \\ \midrule
        d. & RRF -- No Expansion + Entity Text                & \textbf{0.9338}\dag & \textbf{0.8436}\dag & \textbf{0.8124}\dag & \textbf{0.7500}\dag & 0.2372 & 0.0385 & 0.0163 & 0.0019 \\
        e. & RRF --  No Expansion + Entity Hash                & 0.9250\dag & 0.8260\dag & 0.7921\dag & 0.7205\dag & 0.2378 & 0.0367 & 0.0152 & 0.0034 \\
        f. & RRF -- Entity Text \& Hash                 & 0.9231 & 0.8260\dag & 0.7982\dag & 0.7314\dag & 0.2218 & 0.0375 & 0.0161 & 0.0053 \\
        g. & RRF -- No Expansion + Entity Text \& Hash  & 0.9313\dag & 0.8370\dag & 0.8043\dag & 0.7376\dag & 0.2358 & \textbf{0.0391} & 0.0156 & 0.0035 \\
        \bottomrule 
    \end{tabular}
    \label{tab:results-table}
\end{table*}

\subsection{Methods}

\paragraph{BM25 expansion.} We experimented with three retrieval methods to show the benefits of entity annotation for passage ranking: one baseline method and two methods that use query entity expansion~\citep{Shehata} using REL:

\begin{itemize}[leftmargin=*]
    \item[\textbf{a}] \textbf{BM25 -- No Expansion.} As a baseline method, we used BM25 as implemented in Anserini~\cite{Kamphuis2020BM25} using hyper-parameters $k_1=0.82$ and $b=0.68$, shown to be optimal for the MS MARCO dataset. MS MARCO was indexed normally, and no expansion was considered for the queries or the passages. 
    \item[\textbf{b}] \textbf{BM25 -- Entity Text Expansion.} In this method, passages and queries are expanded with the text representation of their annotated entities (from REL). Once the passages and queries have been expanded with entities, we run BM25 with the same hyper-parameter settings as described in \textbf{a}.
    \item[\textbf{c}] \textbf{BM25 -- Entity Hash Expansion.} Instead of using the text representation of entities as an expansion, we expanded the passages and queries by the MD5 hash of the entity text (from REL). The use of MD5 hashing is to provide a consistent representation of multi-word terms and to avoid partial or incorrect matching between queries and non-relevant passages; e.g., passages that contain the word ``united'', do not benefit if the query contains ``United States'' as an entity. Again, after expansion, we run BM25 with the same hyper-parameter settings described in \textbf{a}.
\end{itemize}
In these experiments, the identified entities are deduplicated. As a demonstration of the proposed text expansion methods, Figure~\ref{fig:exp_queries} shows how the query expansion is performed using explicit and hashed forms. The added entities provide more precise context and help eliminate ambiguous terms. Figure~\ref{fig:exp_passage} shows the expansion methods on the relevant passage for this query. The relevant passage can be found through our expansion technique. The linking system recognizes that both the query and the passage contain a reference to the entity \emph{Sacagawea}, even though they are spelled differently in the query and the passage.

\paragraph{Reciprocal Rank Fusion.} As a second series of experiments, we applied Reciprocal Rank Fusion (RRF)~\citep{10.1145/1571941.1572114} to the runs described above. RRF is a fusion technique that can combine rankings produced by different systems. RRF creates a new ranking by only considering the rank of a document in the input. Given a set of documents $D$ and a set of rankings $R$, RRF can be computed as: 
\begin{equation}
    RRF(d \in D) = \sum_{r\in R}\frac{1}{k + r(d)}
\end{equation}
Here $k$ is a hyperparameter that can be optimized, but we simply used a default value of $k=60$ for all settings.

This provides us with four new rankings; the RRF of the pairwise combinations of the three rankings described above and the RRF of all three of these runs:
\begin{itemize}[leftmargin=*]
    \item[\textbf{d.}] \textbf{RRF -- No Expansion + Entity Text.} RRF fusion of runs \textbf{a} and \textbf{b}. The run with no expansions and the run with entity text expansions are considered.
    \item[\textbf{e.}] \textbf{RRF -- No Expansion + Entity Hash.} RRF fusion of runs \textbf{a} and \textbf{c}. The run with no expansions and the run with entity hash expansions are considered.
    \item[\textbf{f.}] \textbf{RRF -- Entity Text + Entity Hash.}  RRF fusion of runs \textbf{b} and \textbf{c}. The run with entity text expansions and the run with entity hash expansions are considered.
    \item[\textbf{g.}] \textbf{RRF -- No Expansion + Entity Text + Entity Hash.} RRF fusion of runs \textbf{a}, \textbf{b}, and \textbf{c}. All three runs are considered. 
\end{itemize}

\subsection{Experimental Setup}

In our experiments, we use MMEAD as a resource to expand queries and passages with entities. The experiments are performed using the MS MARCO v1 passage ranking collection, where only queries containing at least one entity annotation are used. We do not expect meaningful differences for queries without any linked entities, as the expanded query is identical to the original query in that case (due to the simplicity of the method applied here). 

As we expect the linked entities to provide additional semantic information about the queries and passages, we conduct further testing on the obstinate query sets of the MS MARCO Chameleons~\citep{chameleons}, which consist of challenging queries from the original MS MARCO passage dataset. In general, ranking methods show poor effectiveness in finding relevant matches for these queries. Our testing focuses on the bottom 50\% of the worst-performing queries from the subsets of Veiled Chameleon (Hard), Pygmy Chameleon (Harder), and Lesser Chameleon (Hardest), which represent increasing levels of difficulty. 

This gives us four query sets on which we evaluate; (1) all queries that contain entity annotations (\emph{dev} -- 1984 queries), (2) all queries in the hard subset that contain entity annotations (\emph{hard} -- 680 queries), (3) all queries in the harder subset that contain entity annotations (\emph{harder} -- 493 queries), and lastly, (4) all queries in the hardest subset that have entity annotations (\emph{hardest} -- 322 queries).

The experiments are evaluated using Mean Reciprocal Rank (MRR) at rank ten and Recall (R) at rank one thousand. MRR@10 is the official metric for the MS MARCO passage ranking task, while R@1000 gives an upper limit on how well re-ranking systems could perform. The Anserini~\citep{10.1145/3239571} toolkit is used to generate our experiments. 

\begin{figure*}[!t]
    \centering
    \includegraphics[width=.9\textwidth]{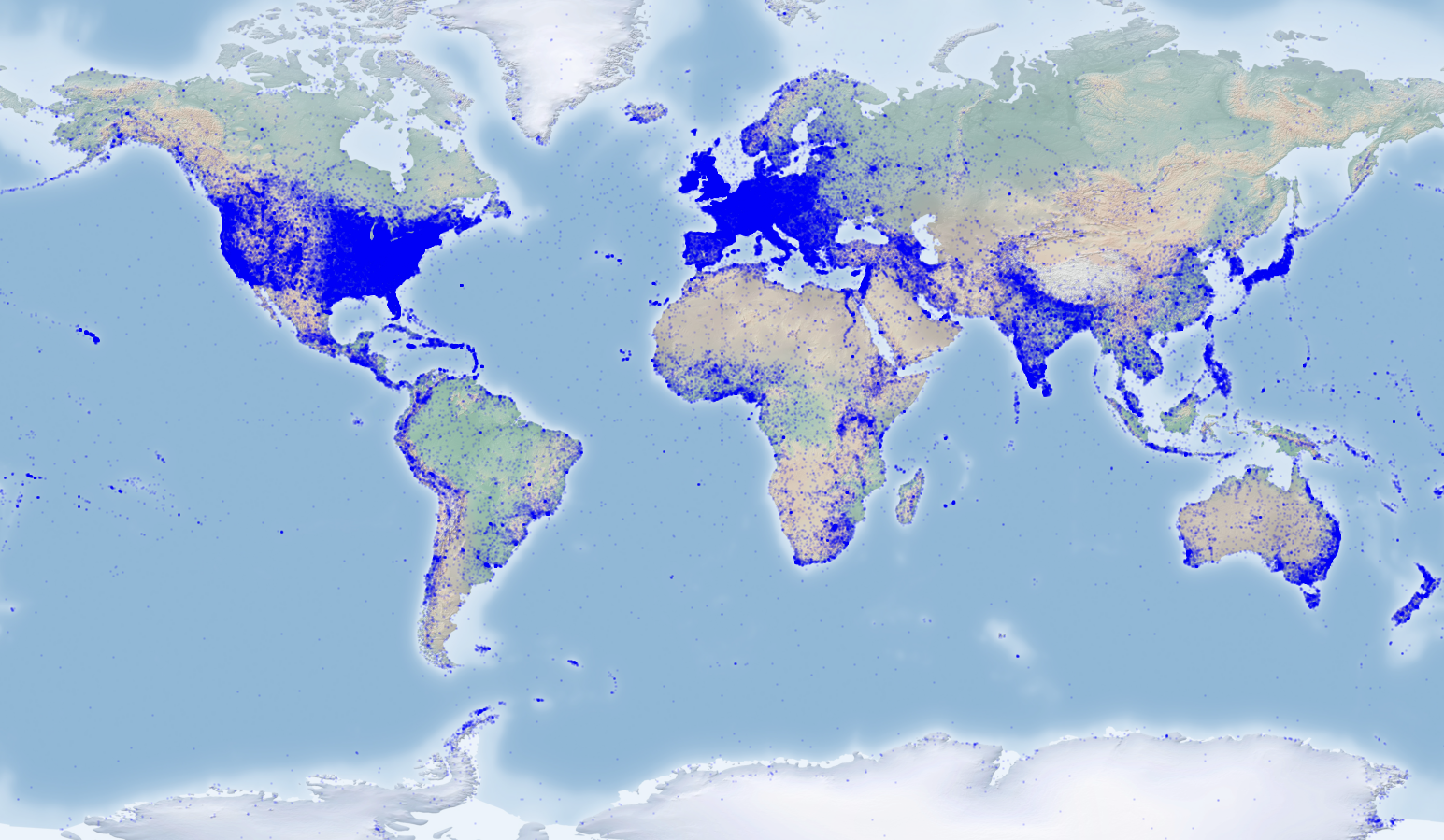}
    \caption{Locations of entities found in the MS MARCO v2 passage collection.}
    \label{fig:entity_map}
\end{figure*}

\begin{figure}[!t]
    \framedtext{
    \begin{itemize}
        \item[\textbf{a.}] did sacajawea cross the pacific ocean with lewis and clark
        \item[\textbf{b.}] \underline{The same text as shown in \textbf{a.}} + \emph{Sacagawea} \emph{Clark} \emph{Pacific Ocean} \emph{C. S. Lewis}
        \item[\textbf{c.}] \underline{The same text as shown in \textbf{a.}} + \texttt{860324} \texttt{a97fed} \texttt{3e3b0e} \texttt{3fe907}
    \end{itemize}
    }
    \caption{Examples of queries for the three different experiments; (\textbf{a}) the non-expanded query, (\textbf{b}) the query with entity text expansion, and (\textbf{c}) the query with entity hash expansion. Text expansions are shown in italics. The MD5 hashes shown in (\textbf{c}) are shortened in this example for formatting.}
    \label{fig:exp_queries}
\end{figure}

\begin{figure}[!t]
    \framedtext{
    \begin{itemize}
        \item[\textbf{a.}] Introduction. Sacagawea, as everyone knows, was the young Indian woman who, along with her baby, traveled with Lewis and Clark to the Pacific Ocean and back.She was a great help to the expedition and many organizations are preparing celebrations to commemorate the 200-year anniversary of the endeavor.y: M. R. Hansen. Sacagawea, as everyone knows, was the young Indian woman who, along with her baby, traveled with Lewis and Clark to the Pacific Ocean and back.
        \item[\textbf{b.}] \underline{The same text as shown in \textbf{a.}} + \emph{Indian Ocean} \emph{James Hansen} \emph{Sacagawea} \emph{India} \emph{Oceania} \emph{William Clark} \emph{Meriwether Lewis} \emph{Pacific Ocean}
        \item[\textbf{c.}] \underline{The same text as shown in \textbf{a.}} + \texttt{fe6fc8} \texttt{860324} \texttt{aa84e6} \texttt{7847ef} \texttt{3e3b0e} \texttt{7d31e0} \texttt{2d8836} \texttt{e58bef}
    \end{itemize}
    }
    \caption{The relevant passage for the query presented in Figure~\ref{fig:exp_queries}; (\textbf{a}) the non-expanded passage, (\textbf{b}) the passage with entity text expansion, and (\textbf{c}) the passage with entity hash expansion. Text expansions are in italics. The MD5 hashes shown in (\textbf{c}) are shortened in this example for formatting.}
    \label{fig:exp_passage}
\end{figure}

\subsection{Results}

Table~\ref{tab:results-table} presents the results of our experiments. If we first look at lines \textbf{a-c} in the results table, we can examine the effects of our expansion methods compared to the baseline run. Looking at R@1000, we can see that more relevant passages are found using entity expansion for the \emph{dev} collection and its harder subsets. We do not find additional relevant documents/passages on the \emph{dev} set when we use the entity hashes, and entity text seems to be the better approach. There is, however, no increase in MRR@10 when using this expansion method. Entity expansions help when evaluating using R@1000, especially when the queries are more complex. The difference in recall effectiveness becomes larger the more complex the queries get. MRR@10 only improves when using entity text expansion.

The reciprocal rank fusion methods are presented in lines \textbf{d-g}. When using these methods, the R@1000 increases more. Again, the subsets that contain more complex queries tend to benefit more. Regarding R@1000 effectiveness, the best RRF method uses a ranking from the normal, not expanded index, with the index that has been expanded with the entity text. Again, entity text expansion helps recall more than using hash expansion. Although the RRF methods improve recall, MRR@10 does not benefit from RRF when compared to using only one of the expansion techniques. 

\section{Beyond Quantitative Results}

In the previous section, we demonstrated the potential value of MMEAD using quantitative evaluations, where we leverage entities to improve retrieval effectiveness in standard benchmark datasets.
Beyond these quantitative results, MMEAD can also help enrich interactive search applications in various ways.
This section describes a few such examples.

Entity links to Wikidata provide an entr\'ee into the broader world of open-linked data, which enables integration with other existing resources.
This allows us to build interesting ``mashups'' or support search beyond simple keyword queries.
As a simple example, we can take the entities referenced in MS MARCO, look up the coordinates for geographic entities, and plot them on a map. 
Figure \ref{fig:entity_map} shows a world map with all entities found in the MS MARCO v2 passage collection mapped onto it (each shown with a transparent blue dot).
The results are as expected, where the blue dots' density largely mirrors worldwide population density, although (also as expected) we observe more representation from entities in North America, Europe, and other better-developed parts of the world.

Figure \ref{fig:entity_map} is a static visualization, but we can take the same underlying data and principles to create interesting interactive demonstrations.
Geo-based search is an obvious idea, where users can specify a geographic region -- either by dragging a box in an interactive interface to encompass a region of interest, or specifying a geographic entity.
For example, the user might ask ``Show me content about tourist sites in Paris''\ and receive passages about the Eiffel Tower in which Paris is not mentioned explicitly.
Simple reasoning based on geographic containment relationships on open-linked data resources would be sufficient for answering this query.
While it is possible that pretrained transformers might implicitly contain this information, they can never offer the same degree of fine-grained control provided by explicit entity linking.

As a simple demonstration, we have taken MMEAD, reformatted the entity links into RDF, and ingested the results into the QLever SPARQL engine~\citep{qlever}.\footnote{\url{https://github.com/ad-freiburg/qlever}, last accessed April 26th 2023}
By combining MMEAD with RDF data from Wikidata and OpenStreetMap, we can issue SPARQL queries such as ``Show me all passages in MS MARCO about France''.

\begin{figure}
\centering
\begin{lstlisting}
PREFIX rdf: <http://www.w3.org/1999/02/22-rdf-syntax-ns#>
PREFIX ex: <http://example.org/> 
PREFIX schema: <https://schema.org/>
PREFIX rdfs: <http://www.w3.org/2000/01/rdf-schema#>
PREFIX passage: <http://example.org/passage> 
PREFIX geo: <http://www.opengis.net/ont/geosparql#>
PREFIX wd: <http://www.wikidata.org/entity/>
PREFIX wdt: <http://www.wikidata.org/prop/direct/>
SELECT ?pid ?content ?entity ?label ?coord 
WHERE {
    ?pid rdf:type passage: .
    ?pid schema:description ?content .
    ?pid passage:has ?entity .
    FILTER (regex(?entity, "wikidata", "i"))
    ?entity rdfs:label ?label .
    ?entity wdt:P625 ?coord .
    ?entity wdt:P17 wd:Q142 .
    FILTER (LANG(?label) = "en")
}
\end{lstlisting}
    \caption{SPARQL query that produces all entities in the passages of the MS MARCO v2 collection that are related to the country of France.}
    \label{fig:code_sparql}
\end{figure}

\begin{figure}[!t]
    \centering
    \includegraphics[width=0.45\textwidth]{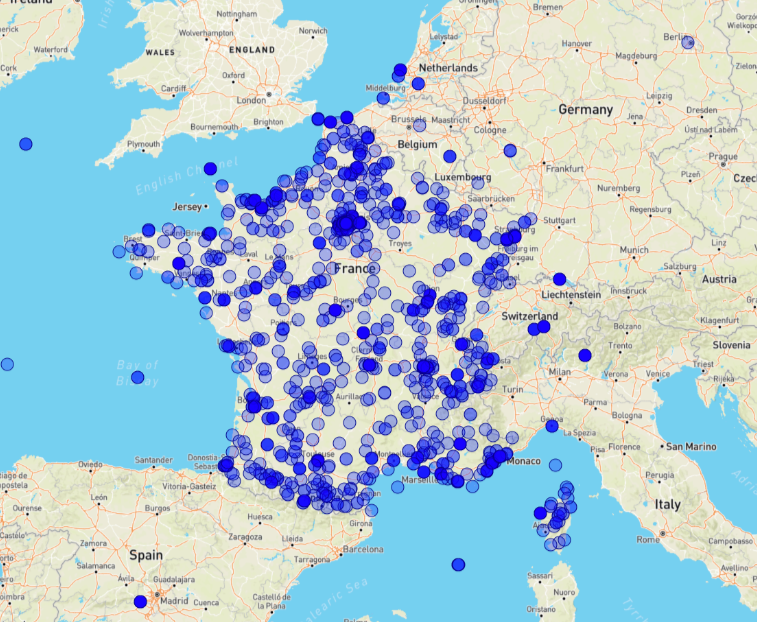}
    \caption{First 1000 entities found in that are connected to France. Entities are represented with a blue dot on the map.}
    \label{fig:france}
\end{figure}

The query is shown in Figure~\ref{fig:code_sparql}, which gives us 122,316 entities found in the collection that have a connection with France (most of them are located in France). Then we can automatically show the entities on a map, as presented in Figure~\ref{fig:france} (showing the first 1000 entities found). 

Not all linked entities are located in France, however. For example, some entities are related to France (entities for which France is mentioned in their Wikidata), but are located elsewhere in the world. One of the blue dots in Germany is the source of the river \emph{Moselle}. This river starts in Germany by splitting off from the \emph{Rhine}, and then goes through France. 
Instead of querying for France, we can also query for different countries. Table~\ref{tab:country_entities} shows the number of entities found for a sample of countries.

\begin{table}[]
    \caption{Number of entities found per country for some example countries where the entity has an English label.}
    \label{tab:country_entities}
    \centering
    \begin{tabular}{l|c|r}
        \toprule
        Country & WikiData ID & \# Entities\\
        \midrule
        United States & Q30 & 3,429,889 \\
        Canada & Q16 & 170,833 \\
        France & Q146 & 122,316 \\
        New Zealand & Q664 & 19,094\\
        Peru & Q419 & 16,448 \\
        Iran & Q794 & 13,633\\
        Ecuador & Q736 & 10,588 \\
        South Korea & Q884 & 9,718\\
        Monaco & Q235 & 8,546\\
        Singapore & Q334 & 6,597\\
        \bottomrule
    \end{tabular}
\end{table}

\section{Conclusion and Future Work}
This research presents the resource MMEAD, or MS MARCO Entity Annotations and Disambiguations. MMEAD contains entity annotations for the passages and documents in MS MARCO v1 and v2. These annotations simplify entity-oriented research on the MS MARCO collections. Links have been provided using the REL and BLINK entity linking systems. Using DuckDB, the data can quickly be queried, making the resource easy to use. 
We also demonstrated that our resource can enrich interactive search applications. In particular, we present an interactive demo where all entities related to geographical locations can be positioned on a map. We experimentally show that MMEAD improves recall effectiveness significantly when using entities for query and passage expansion. When using reciprocal rank fusion, the effectiveness difference becomes even more prominent and new relevant passages are found. The question remains whether these passages can be ranked higher by new retrieval models. With MMEAD, we support information retrieval research that combines deep learning and entity information. 

In the future, we would like annotations from a more diverse group of linking systems. Using the MMEAD format, releasing entity links for collections beyond MS MARCO is also possible. We already showed that using entity links improves recall when using the linked entities for query expansion. What the effects are when training, e.g., DPR methods that include the entity links, is yet to be investigated -- an exciting research opportunity that lies ahead. 

\begin{acks}
This work is part of the research program Commit2Data with project number 628.011.001 (SQIREL-GRAPHS), which is (partly) financed by the Netherlands Organisation for Scientific Research (NWO), and has also been funded in part by the EU project OpenWebSearch.eu under GA 101070014.
Additional support comes from the Global Water Futures program funded by the Canada First Research Excellence Fund (CFREF) and the Natural Sciences and Engineering Research Council (NSERC) of Canada.
\end{acks}

\bibliographystyle{ACM-Reference-Format}
\balance
\bibliography{refs}

\end{document}